# Mechanical Properties of Au Coated Si Nanowafer: an Atomistic Study

Md. Shahriar Nahian[1, a)], Shahriar Muhammad Nahid[2, b)], Mohammad Motalab[1,c)] and Pritom Bose[1, d)]

[1]*Department of Mechanical Engineering, Bangladesh University of Engineering and Technology*

[2]*Department of Mechanical Science and Engineering, University of Illinois at Urbana-Champaign, Urbana, IL 61801, United States*

[a)]shahriarnahian2015@gmail.com

[b)]shahriar.nahid007@gmail.com

[c)] Corresponding Author: mtipuz@yahoo.com

[d)] bose.buet@gmail.com

## 1. ABSTRACT

Combined gold and silicon nano-system has spurred tremendous interest in the scientific community due to its application in different metal-semiconductor electronic devices and solar driven water splitting cells. Silicon, fabricated on gold layer, is prone to gold atom diffusion at its surface. In this study, detailed analysis of mechanical properties of gold coated silicon nanowafer is studied by performing molecular dynamics tensile and compressive simulations. The effects of temperature, gold coating thickness, strain rate and crystallographic orientation of silicon on the mechanical properties are observed for the nanowafer. It is found that both the ultimate tensile and compressive strength show inverse relationship with temperature. The nanowafer fails mainly by slipping along {110} plane due to excessive shear when loaded in [100] direction while a mixed slip and crack type failure occurs for 300K. Interesting crystallographic transformation from fcc to hcp crystal is observed in gold layer for the highest gold layer thickness during tension. The effects of strain rate in tension and compression is also studied. Finally, the crystal orientation of silicon is varied and the tension-compression asymmetry inn the gold coated silicon nanowafer is investigated. Reverse tension-compression asymmetry is observed in case of loading along [110] crystal orientation. The failure mechanism reveals that interesting crystal transformation of silicon occurs during compression leading to early yielding of the material.

## 2. INTRODUCTION

Heteroepitaxial metal-semiconductor (M-S) systems hold sheer interest among the research community, specially due to their wide applicability both as rectifier devices and ohmic contacts depending on the Schottkey Barrier height.[1] They possess interesting electrical [2–4], magnetic [5,6] and various solid state optoelectronic properties [7,8], often derived from Quantum Size effects near such nanocontacts [9].

Silicon is ubiquitously present in the current practice of micro-nano scale nanoscale transistor [10–12] sensors [13] actuators [14], power sources [15] and battery anodes [16] and deems even more possibilities in near future. Metallic nanowires and nanofilms have broad application in different areas such as nano-mechanical and nano-electrical devices. Metal thin films show localized surface plasmon resonance (LSPR) where free electrons collectively oscillate in response to the incident EM wave. Very recently, synergistic devices have been proposed to fully exploit the unique optical, electronic and mechanical properties of both metal and semiconductor nanocomponents. It has been shown that surface plasmonic effect of thin Au or Ag coatings can enhance the amount of light trapped in both thin-film and wafer based Silicon solar cells by 7 to 18 [17–19] folds depending on the wavelength. On the other hand, volumetric fraction of Si shell over Au nanoparticle can modulate Au's plasmonic resonance [20] and such tunability has wide potential in sensor technologies [21,22] and optoelectronic devices [23,24]. Au capped Si nanowires have been reported to show photo detecting capability without being visible to a broad range of EM frequency [25] and Au coated porous Si anodes have been proposed to tackle the challenging lithiation swelling problem [26], making such heteroepitaxial system appealing to a wide variety of fields.

Kang and Cai performed molecular dynamics simulations to investigate the size and temperature effect on the mechanical properties and fracture mechanism of the Si nanowires (NW). [27] Experiments and molecular modeling showed both ductile and brittle failure modes at different conditions. [28,29] Atomistic mechanisms and dynamics of adhesion, nano-indentation, and fracture of Au NWs were investigated by Landman with molecular dynamics simulation and atomic force microscopy [30]. The effects of surface stress driven reorientations was studied extensively by Diao et al. [31,32]. Park presented numerical simulations of gold nano wires under tensile loading at various strain rates and wire sizes at room temperature. Park reported yield strength values ranges from 2.4GPa to 4GPa depending on strain rates. They also observed an increasing trend in the values of yield strength with strain rate [33].

Though different studies have been conducted regarding the mechanical behavior of gold and silicon individually at nanoscale, mechanical properties of gold-silicon combined system are yet to be thoroughly investigated. Since relative orientation between layers often dictates the mechanical response and failure mechanism, orientation dependent behavior of Si-Au system also deserves to be investigated. In this paper we perform molecular dynamics (MD) study for Au coated Si nanowafer and vary the orientation of Si layer with respect to the loading axis. We also observe the effect of loading condition (tension and compression) and explain the asymmetric tensile and compressive properties.

### 3. METHODOLOGY

A Silicon nanowafer (NWa) is generated with the dimensions of 14 nm, 6 nm and 7 nm along the length, width and thickness direction. Two types of Silicon NWa with [111] and [110] crystal directions oriented along the length-axis are prepared to study the effect of crystal orientation with respect to the loading axis. Then a layer of [111] gold coating is added on either side of the silicon wafer. The Silicon and Gold layers' in-plane dimensions are chosen for minimum lattice mismatch and edge strain. In all the simulations, the thickness of gold coating is used as 1.46 nm unless otherwise specified.

Molecular Dynamics simulations are performed to investigate the mechanical properties and fracture behavior of Silicon-Gold heteroepitaxial structures. LAMMPS software package is used for this purpose. Periodic boundary conditions are incorporated along the in-plane directions (x and y). The size of the simulation box was taken to be an exact multiple of a unit cell in both directions so that the effects of free edge site can be suppressed. Enough vacuum is ensured in the out of plane direction to eliminate the effect of interaction between NWa and their periodic images. Conjugate gradient (CG) minimization scheme was used to predict the ground state structure. Then the system was relaxed within a microcanical (NVE) ensemble, followed by isothermal-isobaric (NPT) ensemble at 300 K temperature and atmospheric pressure. Next, a constant strain rate of $10^9 \text{ s}^{-1}$ is applied along the x direction. Such high strain rate simulations are commonly used in nanoscale simulations to investigate different material failure phenomena [27,34–36] and is regarded as one of the scaling issues regarding MD simulations. The strain rate is varied to observe

its effect on the mechanical behavior. Atomic stresses are described by the following virial equation [37]

$$\sigma_{virial}(r) = \frac{1}{\Omega}\sum_i[(-m_i\dot{u}_i \otimes \dot{u}_\iota + \frac{1}{2}\sum_{j \neq i} r_{ij} \otimes f_{ij})] \quad (1)$$

Where the sum is taken for all the atoms in the volume, $m_i$ denotes the mass of atom i, $\dot{u}_i$ is the time derivative of the displacement, $r_{ij}$ denotes the position vector and $f_{ij}$ denotes the interatomic force applied on atom i by atom j.

To validate our approach, we simulated the uniaxial tensile test for pure silicon nanowire having a diameter of 2nm. For calculating the atomic interactions, we used Angular Dependent Potential (ADP) parameters for Au-Si binary system, reported by Starikov et al. [38] Potential formulation of ADP considers pair interaction along with embedding energy and non-central contributions. This potential parameter set have been exploited to simulate the Au adsorption and first-order phase transformation on Si substrate. [39] In this study the Young's modulus and ultimate tensile stress are calculated for 300K and 800K to validate the accuracy of the adopted potential along different temperature regions. We also simulated the uniaxial tensile test for pure gold nanowire. It is to be mentioned that our adopted ADP potential do not give very accurate results for systems containing only gold atoms as indicated in the reference [38]. So, some discrepancies between our obtained results and in the literature for pure gold systems exist. But these discrepancies can be ignored as we are simulating for a system having higher ratio of silicon. The comparison between values obtained by our approach and the ones obtained in the literature is described with the help of the table below-

Table 1: Comparison between values by our present method and existing literature

| Material | Property | Temperature | Present Work | Reference |
|---|---|---|---|---|
| Silicon Nanowire (2 nm diameter) | Ultimate Tensile Stress | 800K | 5 GPa | 5-6.5 GPa [27] |
|  | Young's Modulus | 300K | 113 GPa | 110-115 GPa [27] |
| Gold | Young's Modulus | 300K | 63 GPa | 60-100 GPa [40] |

## 4. RESULTS AND DISCUSSION

In this section, the results obtained by molecular dynamics tensile test simulations are presented in a detailed way. Also, necessary physical explanations and discussions are presented for clear interpretation of the presented result. At first, the stress-strain curves of gold coated silicon nanowafer are obtained for different temperatures and strain rates. The fracture mode and failure mechanism are thoroughly discussed. Then the thickness of the gold coating is varied to realize the effects of gold coating thickness on its mechanical properties. Again, the failure modes and fracture behavior are thoroughly discussed. Moreover, the crystallographic orientation of silicon is varied and mechanical properties for different crystallographic orientations of silicon are analyzed. Finally, the tension compression asymmetry in the mechanical response for different crystallographic orientation is compared and the difference in failure mechanism between compression and tensile test is thoroughly discussed.

## 4.1 Relaxation of Gold Coated Silicon NWa

Initial Au/Si/Au geometry is energy minimized and then sufficiently relaxed under microcanonical ensemble. Microcanonical ensemble (NVE) is chosen because it preserves the system energy maintaining the newton equation in a non-modified form. We observed diffusion of gold particles into the silicon layer as the relaxation is performed (Figure 1). Gold atoms are fast diffusers and they can fill up the pockets, preferentially the interstitial sites, in Si, even in absence of silicon defects. [41] We observed that the diffused gold atoms primarily rest on the silicon interstitial sites. Very few gold atoms were observed to knock silicon atoms to form substitutional defects that indicates a larger energy barrier for silicon substitution. In our simulations, we observed that only atoms from the nearest gold layer of the interfaces diffuse into the silicon layer. Gold atoms were observed to migrate a maximum of about one-layer distance into silicon but created an abruption in the local coordination number to about three layers in gold film adjacent to the silicon interface. The relaxed geometry shows zero residual stress. However, at the onset of deformation, the open gold surfaces start to generate shear bands and later propagate that into the core.

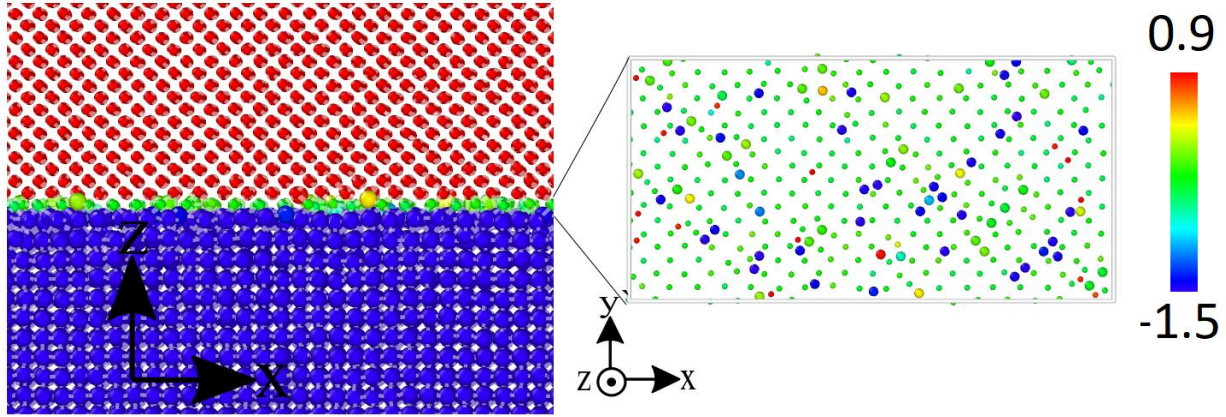

Figure 1: (a) Si/Au interface. Top layer atoms are silicon and bottom layer atoms are Au. (b) close up view of the interfacial layer. Atoms are color coded based on their out of plane distance from the interfacial layer. Au atoms are presented with larger atoms.

### 4.2 Effects of Temperature on Tensile Properties

Mechanical performance of the gold coated nano-wafer during tension is studied under different temperatures. The temperatures are varied from 100K to 500K. It can be seen from Figure 2 that with the increase of temperature, the ultimate strength of the material decreases. This phenomenon is quite expected since at high temperature the atomic movements become more prominent. High atomic movements influence the atoms to overcome the energy barrier more easily and so at fracture strain, the bonds between the atoms get broken almost instantly displaying the brittle type failure. It is also evident that a certain level of noise exists in the curves. The noise refers to the dynamic creation and annihilation of dislocations in the gold layer due to the inclusion of gold atoms in the silicon structure. Also, at high temperatures, diffusion of gold in the silicon structure is more feasible because of the large amplitude of the atomic vibrations resulting in much higher noises at high temperatures.

Figure 2 depicts the effect of temperature on the Young's modulus and ultimate stress of the nanowafer. It is clear from the figure that ultimate stress and young's modulus decreases with the increase of temperature. The Young's moduli were obtained from the stress–strain graphs for the strain < 1% using linear regression. The reason is that with increasing temperature, the oscillations of atoms become more violent and such high level of atomic vibrations create large density of lattice imperfections and dislocations within the material, creating a pathway to the failure point. Again because of higher vibration, the bonds between the atoms become weaker. So, straining the material becomes easier at elevated temperature which in turn leads to lower values of young's modulus. We note that, unlike many 2D systems observed so far [42–44], Young's modulus decreases abruptly with temperature in a non-linear fashion for Au coated Silicon, which indicates.

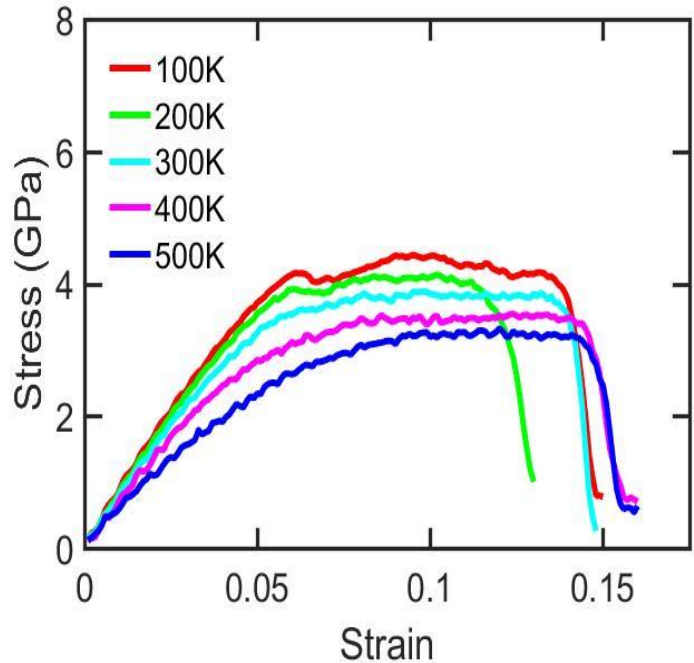

Figure 2: Effects of temperature on the stress-strain response of silicon nanowafer with gold coating thickness of 1.46 nm. The strain rate is selected as $10^9 \, s^{-1}$.

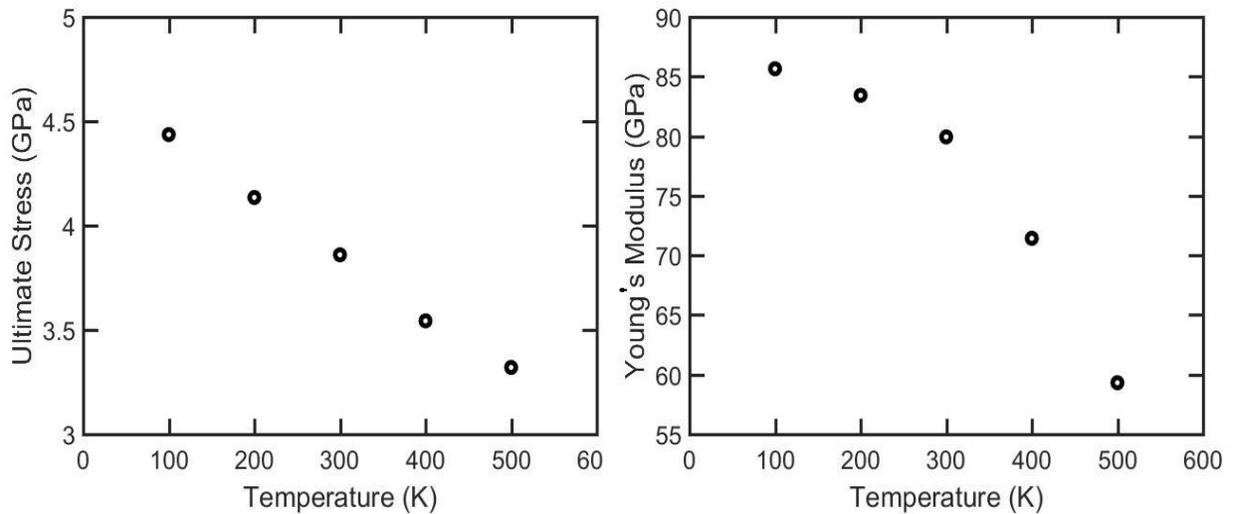

Figure 3: Variation of (a) ultimate stress and (b) Young's modulus of gold coated silicon nanowafer with gold coating thickness of 1.46 nm with the variation of temperature and subjected to tensile loading. The strain rate was fixed at $10^9 \, s^{-1}$.

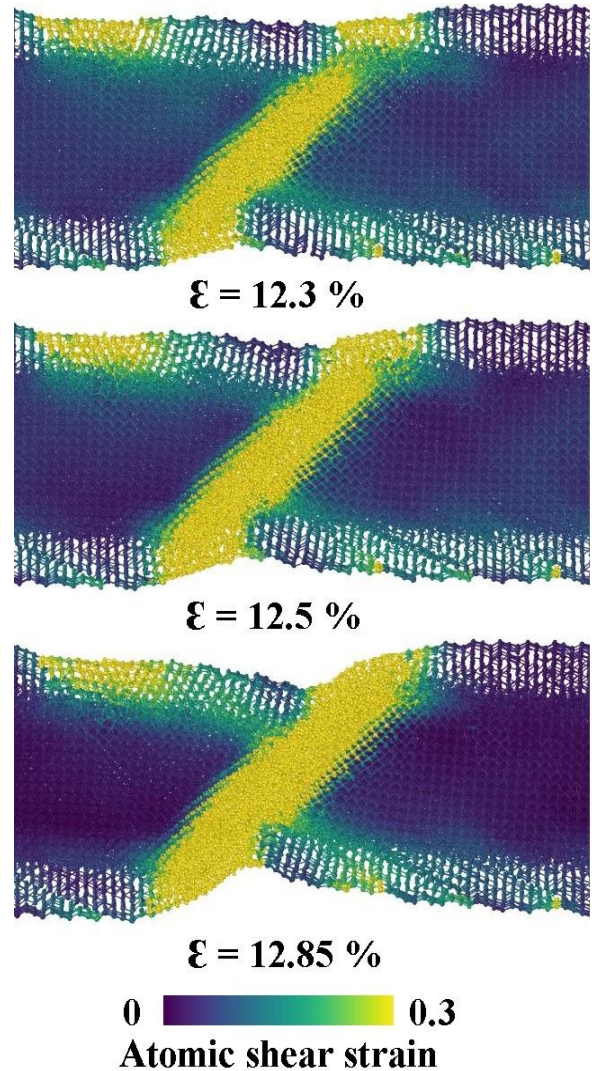

Figure 4: Failure mechanism of gold coated silicon nanowafer at 200K temperature and subjected to tensile loading. Failure initiates by slip due to shear along {111} plane.

## 4.3 Failure Mechanism in Tension for Different Temperatures

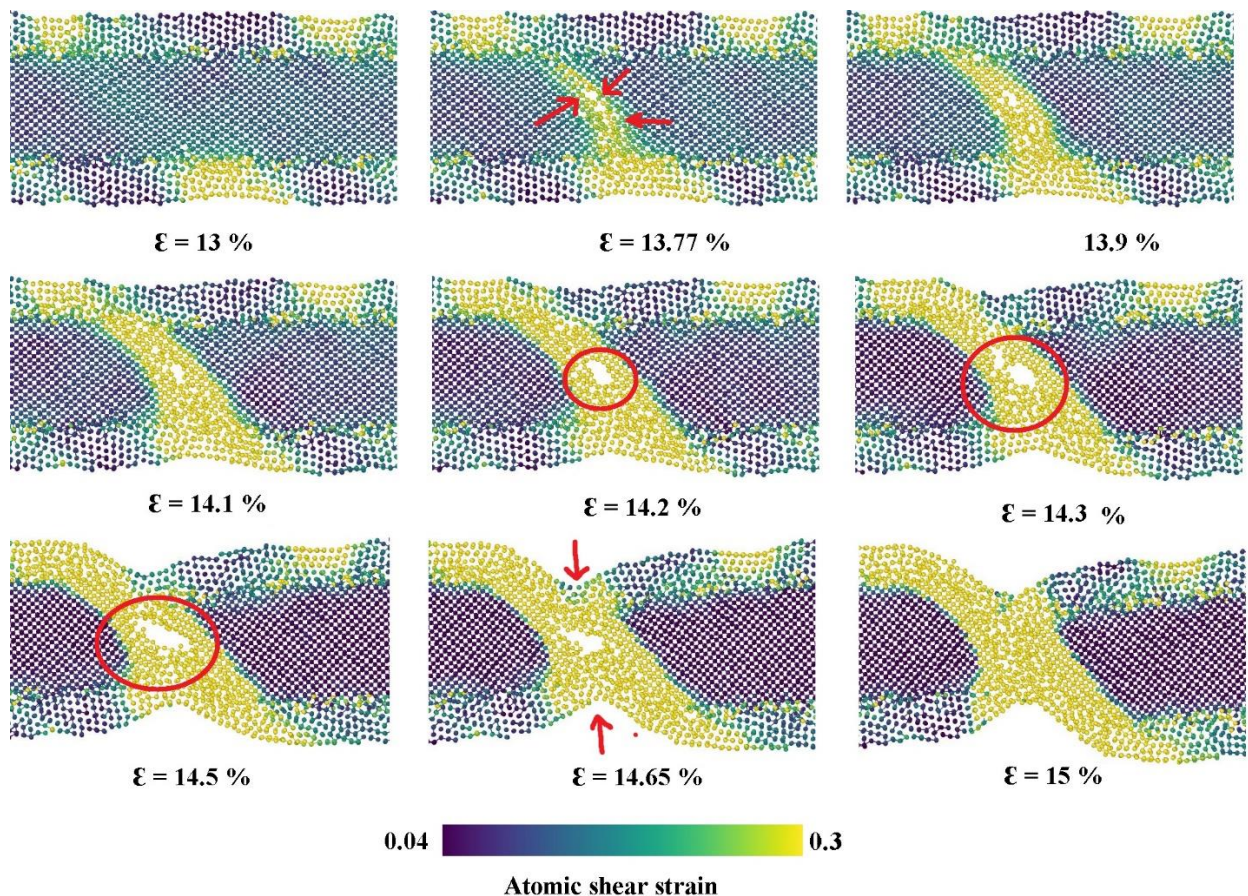

Figure 5: Failure mechanism of gold coated silicon nanowafer at 300K temperature and subjected to tensile loading. Failure initiates by void nucleation and coalescing into crack. Finally, at higher strain, crack propagates and ultimate failure occurs.

In our study, ultimate stress and Young's modulus of the gold coated silicon wafer show inverse relationship with temperature except at 200K where fracture strain a much lower than other temperatures. So a comparison between failure mechanism of silicon – gold system at 200K and 300K is shown in figure 4 and figure 5 respectively. A typical shear failure is observed at 200K which has been discussed earlier. The slip plane is directed along the atoms having much higher atomic shear strain values. At 300K temperature the fracture mechanism is the combination of both crack nucleation and shear failure. Figure 5 shows at 13.7% strain few voids are created in the silicon layer. With further increase in strain values, the crack starts to propagate and necking shows up at the cross section where the crack nucleation and crack propagation occurs. Crack propagation speed can be estimated observing the crack tip position with increasing time steps. Swadener el al determined the maximum crack propagation speed to be 2.59 km/h on the (111) plane in molecular dynamics simulation of silicon beam.[2] Crack propagated in same plane where slip occurs. At 14.65% strain the crack started to disappear and finally failure occurs by shear mechanism.

### 4.4 Effects of Gold Coating Thickness on Tensile Properties

To investigate the effects of gold layer thickness on the mechanical properties of gold coated silicon nanowafer, uniaxial tensile simulation is performed for different thicknesses of gold layer. Figure 3 represents the variation in the stress-strain curves with the variation of gold layer thickness. Five different thicknesses (0.23 nm, 0.78 nm, 1.18 nm, 1.46 nm and 2.27 nm) of gold coating are used while keeping the silicon wafer size constant at 14.1 nm × 6 nm × 3.8 nm. During these simulations we keep the temperature and strain rate constant at 300K and $10^9$ s$^{-1}$ respectively.

From the figure, it can be observed that ultimate stress decreases with the increase of gold layer thickness. The reason is that with the decrease of gold layer thickness, relative amount of silicon increases exerting growingly higher influence on the ultimate strength of the material. Ultimate strength of silicon is much higher than gold. So, with the decrease of gold layer, the material tends to show the behavior of pure silicon and displays higher ultimate strength.

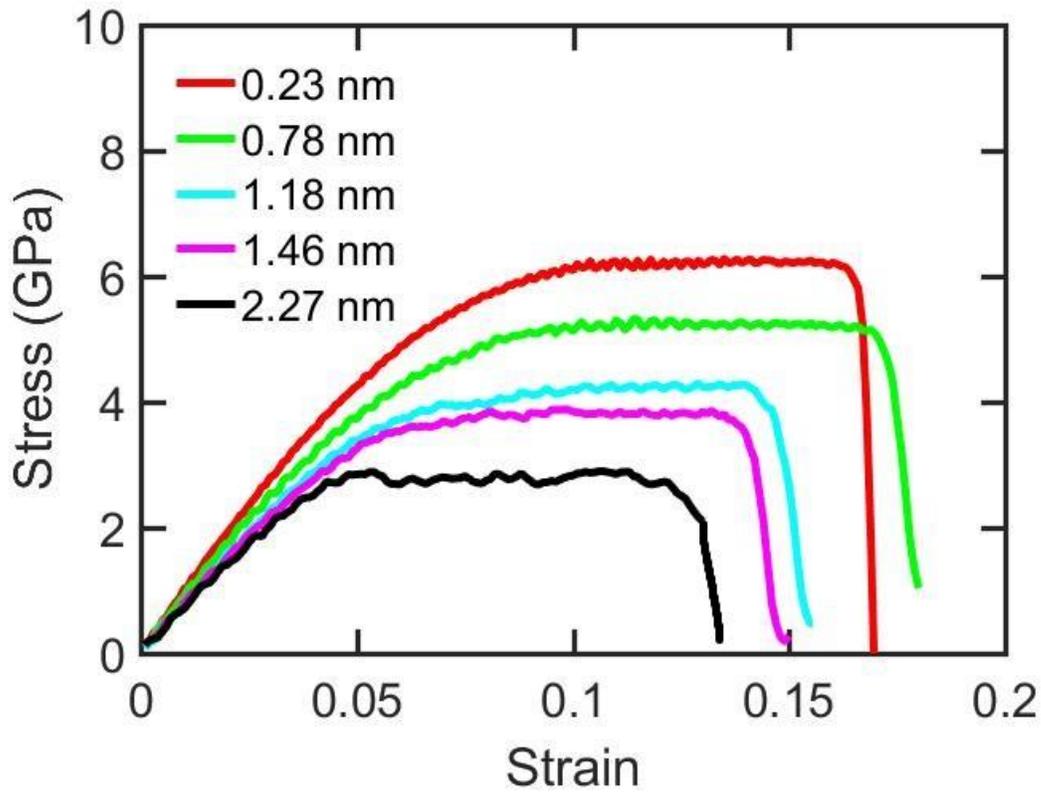

Figure 6: Effects of gold coating thickness on the stress-strain response of gold coated silicon nano-wafer at a temperature 300K. The strain rate is selected as $10^9 \, s^{-1}$.

Existence of a flat portion in the stress-strain curve evident from the figure corresponds to the plastic deformation of the material. The failure of gold coating initiates at around 5% strain. With further increase in the strain, the material shows plastic deformation behavior because of the failure of gold layer. So, it can be concluded that even a tiny portion of gold layer can increase the ductility of silicon. From the figure, we can also observe that nanowafer with 0.78 nm thickness of gold layer shows higher ultimate strength than the one with 0.23 nm thickness. This is because of the fact that nanowafer with 0.78 nm thickness of gold coating has much higher portion of gold in the structure. So, the ductility effect induced by the gold layer for this case is much larger than the one with 0.23 nm thickness of gold layer.

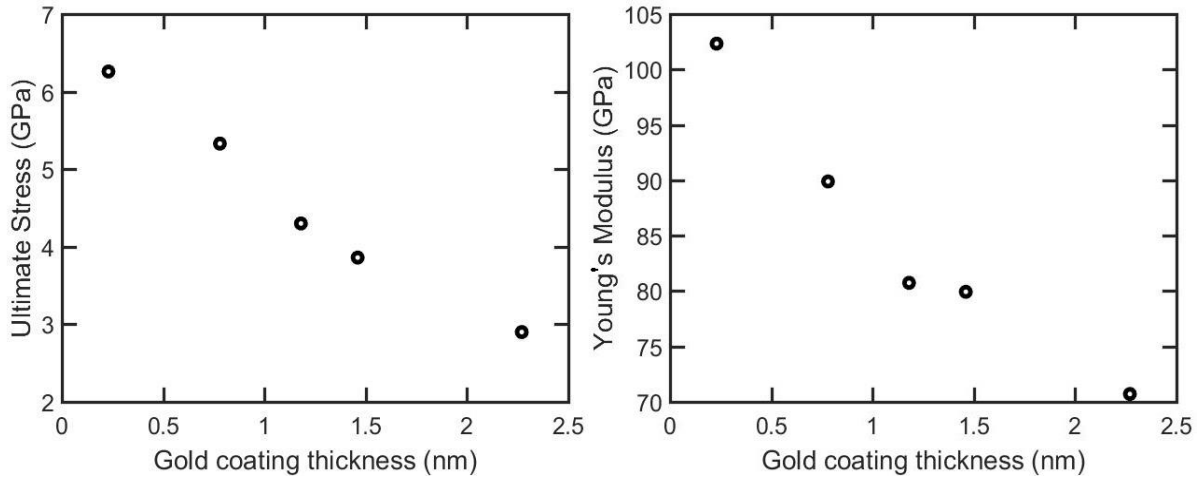

Figure 7: Variation of (a) ultimate stress and (b) Young's modulus of gold coated silicon nanowafer with the variation of gold coating thickness. The temperature and strain rate was fixed at 300K and $10^9$ s-1.

Figure 7 depicts the variation of ultimate stress and Young's modulus of the nanowafer with the variation of gold coating thickness. This study is particularly important to determine the optimum thickness of gold coating. It is evident from the figure that with the increase of gold coating thickness, the ultimate stress and Young's modulus decreases. At the lowest gold coating thickness, the Young's modulus of the nano-wafer is almost equal to the pure silicon. With the increase of gold coating thickness, the amount of gold also increases and thus nanowafer shows lower value of Young's modulus.

### 4.5 Failure Mechanism in Tension for Different Gold Coating Thickness

Almost in all cases the failure mechanism of gold coated silicon wafer is affected by the interfacial phenomena of silicon and gold interface. The inclusion of gold particles in silicon sheet from the interface causes creation of nucleation sites while the wafer is strained. Due to gold inclusion mediated dislocation nucleation sites, gold-silicon system reaches the yield point earlier. Failure of the samples follow the same pattern on the similar account.

Molecular dynamics simulations are conducted varying gold layer thickness for [100] oriented silicon keeping the wafer dimension and temperature constant (300K). The behavior can attributed that almost all the simulated specimen having different coating thickness or relative thickness of silicon layer shows shear failure under uniaxial tension. Figure 8 describes a typical shear failure of silicon wafer having gold layer thickness of 1.18 nm. In this case, extensive sliding on a {101} plane (tilted with respect to loading axis) is observed which causes significant local thinning of nanowafer before fracture. Fracture with shear mechanism refers to a ductile failure. Figure 8 also indicates that initiation of dislocation starts from silicon-gold interface. The movement of dislocation is along the direction perpendicular to slip plane. Displacement vectors are shown to describe the direction of atomic movement at different strains. At 15.2% strain the direction of displacement vectors clearly indicates that the slip plane is along {101} plane.

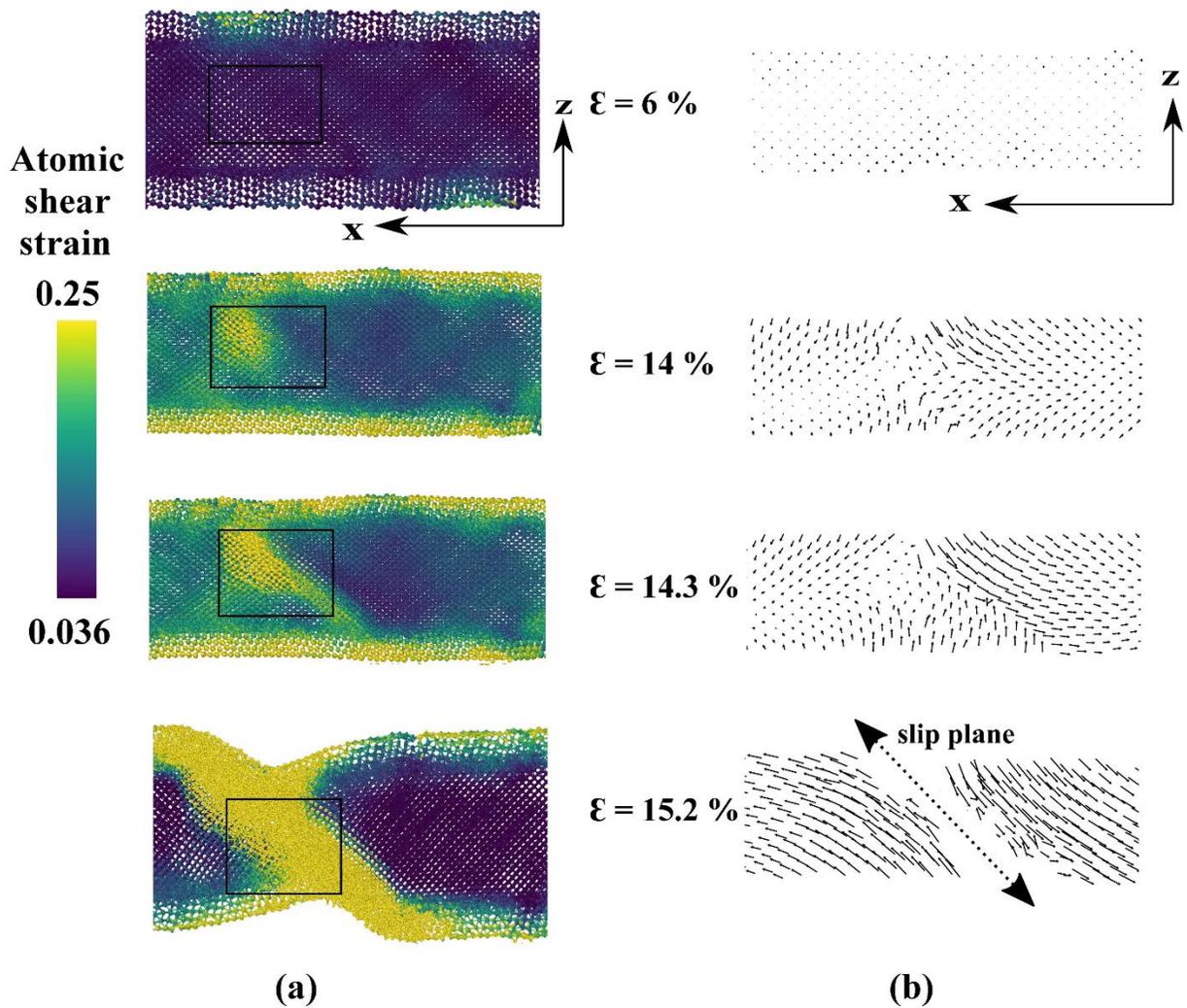

Figure 8 Failure mechanism of gold coated silicon nanowafer for gold coating thickness of 1.18 nm. At different strains. (a) Atomic shear strain along with (b) atomic movement clearly denotes the {110} slip plane.

It is evident from past studies that FCC structures are very prone to be converted into HCP structures under deformation [1-3]. In figure 9 an analogy between atomic shear strain and FCC to HCP transformation is described. Silicon wafer having maximum gold layer thickness of 2.27nm is shown under different strain values. The atomic shear stain value indicates the slip based shear failure in silicon wafer. In this case as the gold layer thickness in higher than the other wafers, the gold layer shows significant effect on the overall behavior of the nanowafer. With the increase in value of longitudinal strain, the percentage of HCP atoms increases. It is clear from the figure 9 that atoms having much higher atomic strain values tend to transformed into HCP structure from FCC. This conversion increases the ductile nature of the overall system.

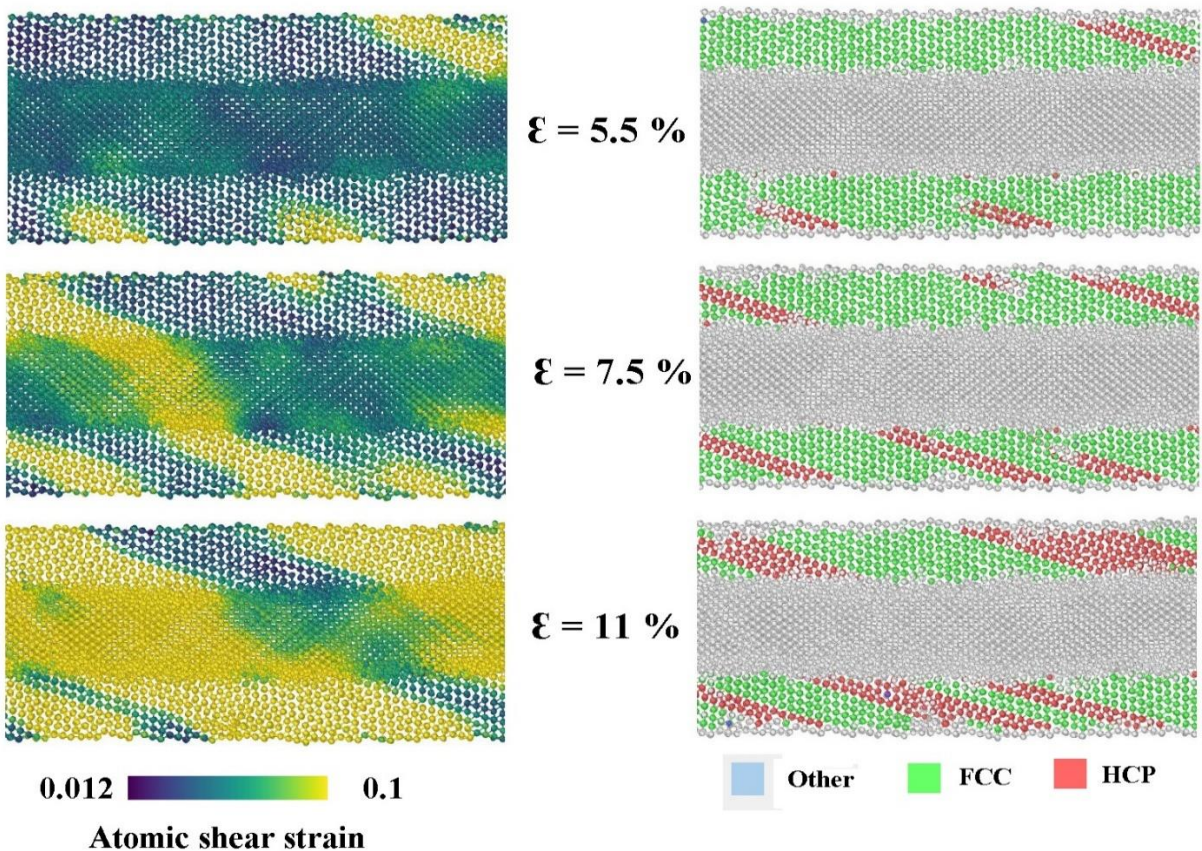

Figure 9 Failure mechanism of gold coated silicon nanowafer for gold coating thickness of 2.27nm.. Crystal transformation from FCC to HCP occurs in gold layer due to shear. Figure (a) shows the atomic shear strain and Figure (b) shows the crystal transformation from FCC to HCP

## 4.6 Effects of Crystallographic Orientation on Tensile Properties

Figure 10 depicts the stress-strain curves of gold coated silicon naowafer for different silicon orientation for tensile loading. Material strength is the highest for [111] orientation and lowest for [100] orientation. Also, it is evident from the figure that, in case of [111] orientation, stress-strain relationship is similar to brittle type material while for [100] orientation, the relationship is ductile type. The difference in stress-strain curve is more elaborately discussed in a future section.

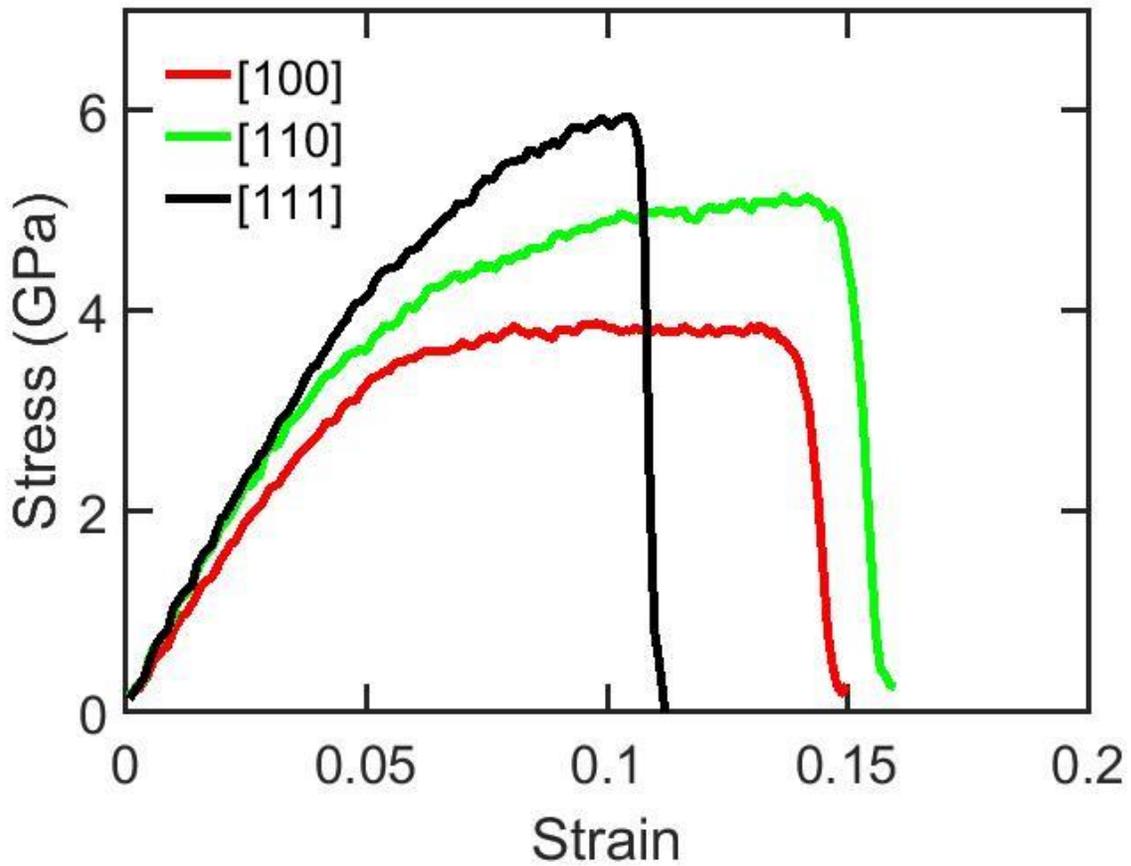

Figure 10 Stress strain relationship of gold coated silicon nanowafer for various different orientations of silicon while subjected to tensile loading. The temperature and strain rate are selected as 300K and $10^9 s^{-1}$

## 4.7 Failure Mechanism in Tension for Different Crystal Orientation of Silicon

Failure of gold coated silicon nanowafer having silicon crystal orientation of [100] towards the length direction occurs due to the void nucleation and crack propagation in the slip plane. This mechanism was discussed in earlier.

Figure 11 describes the failure mechanism of gold coated silicon nanowafer having [110] silicon crystal orientation in length direction. The atomic strain distribution is showed in different strains. From the figure, it is clear that material failure occurs due to shear in {111} plane which lead to slip and cause ultimate failure of the material. This type of failure is called ductile type failure. For [111] silicon orientation, failure mechanism is quite different which is described in figure 12. At high strain (about 10.7 %), crack initiates from the interface of gold and silicon and propagates through the material. The failure initiation and propagation is extremely quick which corresponds to brittle type material behavior. Also, {111} crack plane is observed for this case.

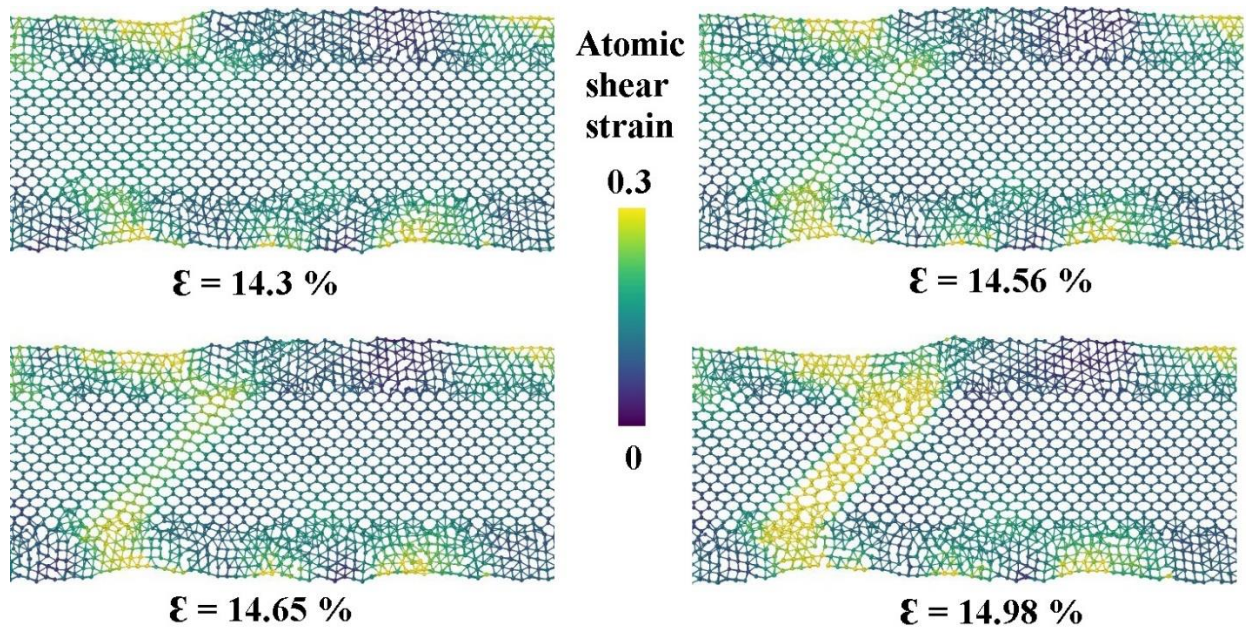

Figure 11 Atomic shear strain distribution depicting the failure mechanism of gold coated silicon nanowafer having [110] silicon crystal orientation in the length direction and subjected to tensile loading. Ductile type failure occurs due to slipping along {111} plane.

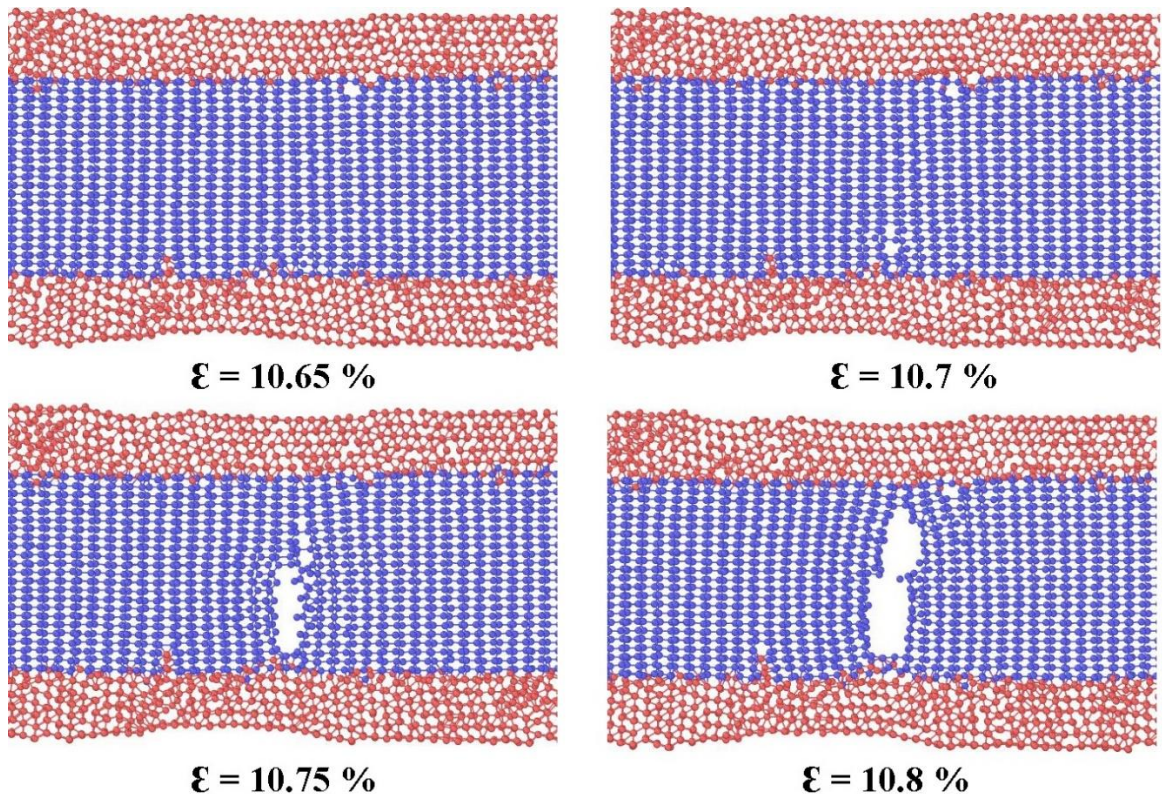

Figure 12 Failure mechanism of gold coated silicon nanowafer having silicon crystal orientation of [111] directed along the length axis. Crack initiates from the gold-silicon interface at about 10.65 % strain which leads to ultimate material failure at about 10.8 % strain.

## 4.8 Tension-Compression Asymmetry for Different Crystal Orientation of Silicon

Studying the tension-compression asymmetry for nanoscale materials have been the focus of many investigations to gain a deeper understanding of mechanical response and failure behavior of the material. In this section, we discuss the tension-compression asymmetry of gold-coated silicon nanowafer for different crystallographic orientations of silicon. Compressive strength is usually higher than tensile strength for different materials. Similar higher compressive strength are observed for silicon nanowafer having [100] and [111] crystal orientations of silicon. However, in case of [110] crystal orientation, reverse tension-compression asymmetry is observed. Stress decrease sharply and then fluctuates up to a high global strain before ultimate failure.

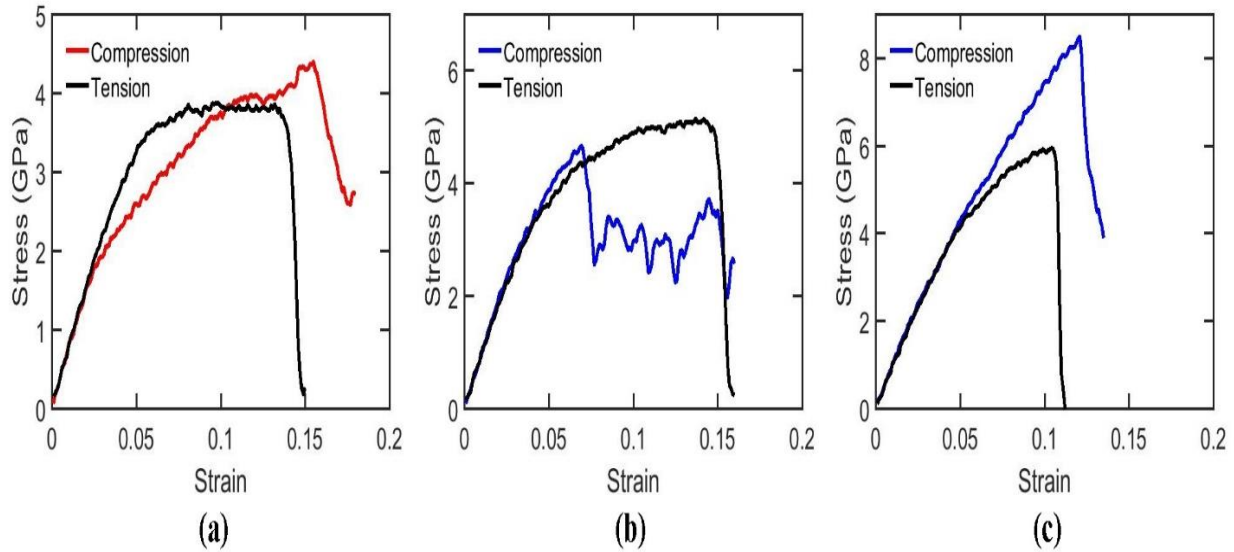

Figure 13 Tension-compression asymmetry for gold coated silicon nanowafer having (a) [100], (b) [110] and (c) [111] crystallographic orientation along length direction.

For [110] silicon orientation, 1/2 <110> dislocation nucleation occurs at about 7.21 % strain as shown in Figure 11 by the evolution of high local atomic shear strain along the {111} plane. The dislocations propagate through the silicon crystals. Silicon crystals having cubic diamond structures transform into hexagonal diamond crystals along with the dislocations movement. This transformation of crystal structure of silicon resists the ultimate failure of the material. At much higher strain, another shear plane along {111} direction is observed. Similar dislocation movements leading to the transformation of silicon crystals from cubic diamond to hexagonal diamond crystals structures are also seen (Figure 14 Finally, failure occurs due to excessive shear along multiple {111} planes at different positions of the material at about 16% strain.

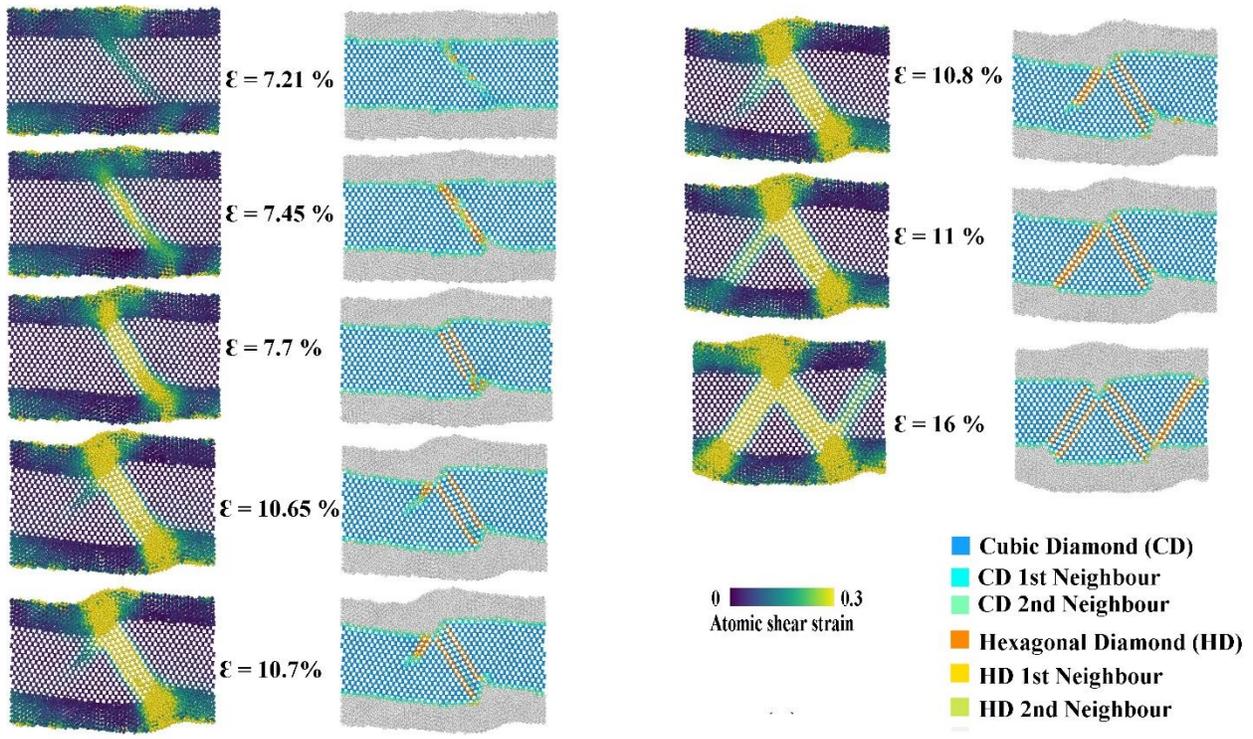

Figure 14 Failure mechanism for gold coated silicon nanowafer having silicon crystal orientation of [110] along the length axis. Atomic shear strain and identification of different diamond structures are depicted at different strains. Cubic diamond structures of silicon are transformed into hexagonal diamond structures due to shear.

## 5. CONCLUSION

Studying the mechanical properties of combined gold and silicon multilayered systems are of extreme importance due to their existing as well as potential applications in various electromechanical nanosystems. In this study, for the first time, extensive molecular dynamics simulation based study is performed to determine different mechanical properties and fracture mechanism of gold coated silicon nanowafer by tensile and compressive test. The effects of temperature, strain rate, gold coating thickness and orientation of silicon crystals are observed. Also, the tension-compression asymmetry is discussed. The following conclusions can be drawn from the results described in the previous chapters:

- Temperature has profound effect on mechanical properties like Ultimate stress, Young's modulus etc under uniaxial tension and compression. Both ultimate stress and Young's modulus decreases with increasing temperature. Gold coated silicon nanowafer having [100] silicon layer orientation exhibits shear and mixed type failure mechanism at different temperatures.

- Silicon -Gold combined system does not generate any synergic effect on the mechanical properties like Ultimate stress and Young's modulus. Interfacial phenomena affect the overall mechanical behavior of the combined system. In maximum cases dislocation or crack nucleation starts from the silicon-gold interface.

- Continuous dislocation creation and annihilation leads to high noise and fluctuations in the stress-strain curves of the material.

- The material shows high strength at high strain rate for both tensile and compressive loading. Young's modulus also shows similar trend with different strain rates.

- Gold layer thickness influences the inclusion of gold particles in silicon layer and vice versa. With increase in value of gold layer thickness ultimate tensile and compressive strength decreases. Almost all simulated wafers having different gold layer thickness experiences ductile type failure (shear failure). Under compressive loading the specimen

having thicker gold layer shows FCC to HCP transformation. 42

- For 2.27 nm thickness of gold coating, crystal transformation of gold atoms from fcc to hcp is observed.
- Loading along three different orientations of silicon atoms are applied to observe the effects of silicon orientation. Loading along [111] direction shows the highest strength and brittle type failure. Failure occurs along {111} plane due to crack creation from goldsilicon interface and propagation through silicon.
- Tension-compression asymmetry for different crystal orientation of silicon atoms are observed. Reverse tension-compression asymmetry is observed for the case of [110] oriented silicon. Silicon phase transformation from cubic diamond to hexagonal diamond occurs due to dislocation propagation through the silicon atoms leading to alteration of crystal structure.